\documentclass[twocolumn,numberedappendix]{aastex62}

\begin{document}

\title{ALICE Data Release: A revaluation of HST-NICMOS coronagraphic images}
\shorttitle{ALICE Data Release Document}
\shortauthors{Hagan et al.}

\author{J. Brendan Hagan}
\affiliation{Space Telescope Science Institute, 3700 San Martin Drive, Baltimore, MD 21218, USA}
\email{hagan@stsci.edu}

\author{\'Elodie Choquet}
\altaffiliation{Hubble Fellow}
\affiliation{Department of Astronomy, California Institute of Technology, 1200 E. California Blvd, Pasadena, CA 91125, USA}
\affiliation{Jet Propulsion Laboratory, California Institute of Technology, 4800 Oak Grove Drive, Pasadena, CA 91109, USA}

\email{echoquet@jpl.nasa.gov}

\author{R\'emi Soummer}
\affiliation{Space Telescope Science Institute, 3700 San Martin Drive, Baltimore, MD 21218, USA}
\email{soummer@stsci.edu}

\author{Arthur Vigan}
\affiliation{Laboratoire d'Astrophysique de Marseille, 38, rue Frédéric Joliot-Curie, 13388 Marseille cedex 13}
\email{arthur.vigan@lam.fr}

\begin{abstract}
The Hubble Space Telescope (HST) NICMOS instrument has been used from 1997 to 2008 to perform coronagraphic observations of about 400 targets. Most of them were part of surveys looking for substellar companions or resolved circumstellar disks to young nearby stars, making the NICMOS coronagraphic archive a valuable database for exoplanets and disks studies. 
As part of the Archival Legacy Investigations of Circumstellar Environments (ALICE) program, we have consistently re-processed a large fraction of the NICMOS coronagrahic archive using advanced PSF subtraction methods. 
{We present here the high-level science products of these re-analyzed data, which we delivered back to the community through the Mikulski Archive for Space Telescopes (MAST) archive:\dataset[10.17909/T9W89V]{http://archive.stsci.edu/doi/resolve/resolve.html?doi=10.17909/T9W89V}.} 
We also present the second version of the HCI-FITS format (for High-Contrast Imaging FITS format), which we developed as a standard format for data exchange of imaging reduced science products. 
These re-analyzed products are openly available for population statistics studies, characterization of specific targets, or detected point source identification.
\end{abstract}



\section{Introduction}

Direct imaging of exoplanetary systems is one of the greatest challenges in modern astronomy due to the very high contrast between the host star and surrounding circumstellar objects at close angular separation. To obtain direct astrophysical signal of these circumstellar objects, dedicated coronagraphic instruments are required to attenuate the starlight in the images to enable fainter object detection. Yet because of imperfect starlight suppression and wavefront variations, the coronagraphic images are still dominated by the star light at short separations. To detect the faintest objects, this residual starlight{, i.e. the coronagraphic point spread function (PSF)\footnote{From hereafter the authors use ‘‘PSF’’ when referring to residual starlight}}, must be subtracted out using post-processing data reduction techniques.

{Classical differential imaging methods consist in subtracting a typical PSF image from the science image, using a single image \citep{Lowrance2005} or a median image \citep{Marois2006} either from a reference star or from the science target itself using angular diversity to limit self-subtraction of the astrophysical source. These classical differential imaging methods usually have limited performance due to significant wavefront variations between acquisitions, causing temporal variations in the PSF. For HST's instruments, instrumental deformations on timescales from a few seconds to several days \citep{Lallo2006,Makidon2006} introduce additional temporal variations in the PSF that typically limit point source detections to contrast lower than $10^4$  within $\sim1\arcsec$ from the star with differential imaging \citep[e.g.][]{Lowrance2005}}. Yet, these methods have been very successful at detecting brown dwarf companions and circumstellar disks \citep[e.g][]{Schneider1999, Kalas2006,Schneider2014}.  

Over the last decade, advanced post-processing techniques have emerged that take advantage of a library of instrument PSFs. {The temporal variations of the PSF introduced in the instrument PSF library can be combined to create an optimal synthetic PSF that is then subtracted from the target PSF.} The synthetic PSF can be created using a linear combination of instrument PSF images (LOCI and its variants \citep{Lafreniere2007}) or using Principal Component Analysis (PCA) on the PSF library \citep{Soummer2012, Amara2012}. For first-generation instruments on the Hubble Space Telescope (HST), these techniques are more efficient when using PSFs from many different stars (Multiple Reference Star Differential Imaging, MRDI), even when observations are separated by large time intervals. This has been demonstrated by the re-discovery of HR 8799 planets (b,c, and d) in archival NICMOS data from 1998 \citep{Lafreniere2009,Soummer2011}. Indeed, depending upon the size and quality of the PSF image library, these techniques can offer a significant improvement in comparison to classical PSF subtraction.

With the advent of these advanced post-processing techniques, it is worth considering reanalyzing some of the data obtained with first-generation coronagraphic instruments. We started the ALICE project (Archival Legacy Investigations of Circumstellar Environments) with the goal to consistently reprocess the NICMOS coronagraphic archive with advanced post-processing methods \citep{Choquet2014d}. NICMOS operated on-board Hubble for about 8 years between 1997 and 2008. Its mid-resolution channel NIC2 (pixel size 0.076”) was equipped with a 0.3”-radius coronagraphic mask and a Lyot stop. During its time in operation approximately 400 stars were observed, most as part of surveys looking for debris disks and planets around nearby stars. The ALICE pipeline assembles and aligns large PSF libraries from consistent subsets of this database that are used to process each individual targets with the KLIP algorithm \citep{Soummer2012}. {This project has revealed new images of 12 debris disks previously undetected from the NICMOS data, among which 11 had never before been imaged in scattered light \citep[][Choquet et al. 2018, in press]{2014ApJ...786L..23S,2016ApJ...817L...2C,Choquet2017}.} In addition, we found a total of 452 point sources uncovered in the data \citep{2015SPIE.9605E..1PC}. 

{We are now delivering the re-processed products of the ALICE program back to the community, openly accessible as High-Level Science Products through the MAST archive\footnote{\href{https://archive.stsci.edu/prepds/alice/}{https://archive.stsci.edu/prepds/alice/}},\dataset[10.17909/T9W89V]{http://archive.stsci.edu/doi/resolve/resolve.html?doi=10.17909/T9W89V}.} We also further improved the HCI-FITS standard format for high-contrast imaging science products to make it compatible with any high-contrast imaging instrument, and we present here version two of this format.

In Sec. 2 we present the content of the ALICE archive and the reprocessed datasets. In Sec. 3 we present the HCI-FITS format used for the delivered science products.

\section{Released datasets}

\subsection{ALICE inputs}
The input data for the ALICE program come from the Legacy Archive PSF Library and Circumstellar Environments (LAPLACE) program\footnote{\href{https://archive.stsci.edu/prepds/laplace}{https://archive.stsci.edu/prepds/laplace}} \citep[HST program AR-11279, PI: G. Schneider;][]{Schneider2010}. This program delivered a homogeneous re-calibration of a large fraction of the raw NICMOS coronagraphic archive, optimized for imaging at separations close to the coronagraphic-mask inner working angle (radius of 0\farcs3) using PSF subtraction techniques. This re-calibration was performed using contemporary flat-field frames optimally matched to the location of the coronagraphic mask (as opposed to epochal flats), and using an improved bad-pixel correction. During the second era of NICMOS operations (after replacement of its cooling system), dark calibration observations were obtained with less frequency than during the first era, and the LAPLACE program delivered two versions of re-calibrated images: one using observation-optimized dark frames when available, and one using synthetic model dark frames. 

For consistency, the ALICE program re-analyzed all the non-polarimetric data from the LAPLACE program that were calibrated with contemporary flats for NICMOS Era 1\footnote{\href{https://archive.stsci.edu/missions/hlsp/laplace/dd1/LAPL/NICMOS-LAPL-DD1/LAPL\_DATA/contemp\_flats/repaired/}{https://archive.stsci.edu/missions/hlsp/laplace/dd1/LAPL/NIC\-MOS-LAPL-DD1/LAPL\_DATA/contemp\_flats/repaired/*\_clc\_*.fits}}, 
and with contemporary flats and observed dark frames for NICMOS Era 2\footnote{\href{https://archive.stsci.edu/missions/hlsp/laplace/dd2/LAPL/NICMOS-LAPL-DD2/LAPL\_DATA\_DD2/comtemp\_flats-DD2/}{https://archive.stsci.edu/missions/hlsp/laplace/dd2/LAPL/NIC\-MOS-LAPL-DD2/LAPL\_DATA\_DD2/comtemp\_flats-DD2/*\_o\_clc\_*.fits}}, which represent 72\% of the non-polarimetric NICMOS archival datasets. Furthermore, we also re-analyzed a few selected datasets that were not re-calibrated by LAPLACE program (from HST programs 7248, 10897, and 11155). 

\subsection{ALICE outputs}

The outputs of the ALICE program -- and delivered data described through this document -- all come from the input data detailed above. Data that was not reprocessed may have: (1) not been part of the LAPLACE inputs described above, (2) suffered from bad acquisition images not centered on the coronagraphic mask, (3) failed sub-pixel alignment within a PSF library. The complete list of NICMOS coronagraphic programs (non-polarimetric data) is detailed in Table~\ref{tab:nicmosPrograms}. The table also reports the number of targets per HST program that have been partly or entirely re-processed as part of the ALICE program. The complete list of re-processed images for each HST program, target, and filter set can be accessed on the MAST archive. Details on how these data were re-analysed (field-of-view, alignment procedure, PSF subtraction and analysis methods) are described in \citet{Choquet2014d}.

NICMOS data from 401 targets were used as input for the ALICE program, some observed in multiple HST programs (451 different target observations) and/or with multiple filters. In total there were 590 different datasets, i.e. 590 unique observations with respect to program, target, and filter. The number of input images that were re-analyzed by the ALICE program amounts to a total of 4879 images (83\% of them were acquired with the F110W or F160W filters). 
Reprocessed data are delivered for 494 datasets out of the 590 input ones. {Of the 96 datasets not delivered, 56 of them were entirely composed of bad acquisition images and 40 others could not be properly aligned within a PSF library (often because the target was not a bright star or the target was a binary or contained a bright companion causing the alignment to fail)}. The majority of these datasets were acquired with the F110W or F160W filters (408 datasets). This amounts to a total of 3955 reprocessed images (86\% in F110W or F160W), delivered back to the community.

\subsection{Note on the reprocessing efficiency}

The PSF libraries assembled by the ALICE pipeline are much larger in the F110W and F160W filters than in the other filters, as the F110W and F160W were commonly used for surveys and specific object characterizations. The medium-band and narrow-band filters were exclusively used for characterization of known circumstellar objects. Furthermore, as ALICE's reprocessing is based on MRDI and excludes images with detected circumstellar material from the reference PSF libraries \citep[except in ``planet mode'', which includes images of the science target acquired in a different orientation of the spacecraft, when available,][]{Choquet2014d}, the medium- and narrow-band filter data were reprocessed with PSF libraries of very small sizes (see Table~\ref{tab:refLib}). ALICE's reprocessing for these datasets is not expected to improve much upon classical PSF subtraction techniques. The real added value from the ALICE program mostly concerns NICMOS data acquired with the F110W and F160W filters.

\begin{deluxetable}{c|rr|rr}
\tabletypesize{\scriptsize}
\tablecaption{ALICE reference PSF Libraries\label{tab:refLib}}
\tablehead{
\colhead{Filter}	& \multicolumn{2}{c}{Era 1} & \multicolumn{2}{c}{Era 1}\\
\cline{2-3} \cline{4-5}
\colhead{}	&\colhead{\# images\tablenotemark{a}}&\colhead{\#  PSF\tablenotemark{b}}&\colhead{\# images\tablenotemark{a}}&\colhead{\#  PSF\tablenotemark{b}}
}
\startdata
\hline
F110W	&	144	&54	& 1854	&809	\\ 
F160W	& 	824	&360	&1248	&655	\\
F165M	& 	42	&16	&49		&19	\\
F171M	& 	36	&16	&	66	&6	\\
F180M	& 	98	&46	&	69	&14	\\
F187N	& 	5	&0	&	9	&6	\\
F187W	& 	3	&0	&	0	&0	\\
F190N	& 	5	&0	&	18	&0	\\
F204M	& 	4	&4	&	74	&19	\\
F205W	& 	11	&0	&	0	&0	\\
F207M	& 	174	&76	&	0	&0	\\
F212N	& 	28	&10	&	0	&0	\\
F215N	& 	28	&10	&	18	&0	\\
F222M	& 	15	&0	&	52	&7	\\
F237M	& 	5	&5	&	0	&0	
\enddata
\tablenotetext{a}{Number of images used as input of the ALICE program.} 
\tablenotetext{b}{Number of images used as references for PSF subtraction.}
\end{deluxetable}


\section{Released products}

For each dataset re-processed by the ALICE program, we provide three kind of outputs: 1) a high-level science FITS file, gathering all the high-contrast imaging metrics in the standard HCI-FITS format for the combined data at the target level. This is the main output product of the ALICE program and should be used for detection purposes. 2) A folder with FITS files for each reprocessed image of the dataset, gathering the material needed to perform forward modeling of an astrophysical signal. These should be used for characterization of signals detected in the main high-level science product. 3) A folder with preview PDF files of the main outputs of the dataset.

\begin{deluxetable*}{llll}
\tabletypesize{\scriptsize}
\tablecaption{Structure of the HCI-FITS files \label{tab:targetLevelFITS}}
\tablehead{
\colhead{Label}&\colhead{Type} &\colhead{Dimensions}& \colhead{Description}
}
\startdata
\hline
	DATA\_INFORMATION	        & BINTABLE	&$N_{im} \times 12$	& Characteristics of each image\\
	REDUCED\_DATA		        & IMAGE		&$N_{im} \times N_x \times N_y$	&Reduced data products\\
	SNR\_MAP			        &IMAGE		&$N_{im} \times N_x \times N_y$	& SNR maps\\
	SENSITIVITY\_MAP 	        &IMAGE		&$N_{im} \times N_x \times N_y$	&2D detection limits\\
 	DETECTION\_LIMIT		    &BINTABLE	&$N_{im} \times 2 \times N_\rho$\tablenotemark{a}	&Radial detection limits\\
	SOURCE\_DETECTION [optional]&BINTABLE	&$N_{source} \times 20 \times N_{im}$	&Detected point sources characteristics
\enddata
\tablenotetext{a}{{$N_\rho$ is number of resolution elements}}
\end{deluxetable*}

\subsection{Definition of dataset}

We refer to a \emph{dataset} as the set of images acquired with the same filter element, as part of the same HST program, and under the same target identifier (HST \verb+TARGNAME+ keyword). Datasets are thus composed of an homogeneous group of images as designed by the PI of the program. A dataset may combine images acquired with different spacecraft orientations, with different exposure times, and different epochs (within a year). We assume that any potential astrophysical source remains unchanged in all the images of a dataset when combining them in the main ALICE science FITS file.
 
A few targets have several datasets with the same filter element, in different HST programs (despite HST policies not to re-observe a target in the same mode), or within the same program (target re-observed after a failed acquisition).

\subsection{The HCI-FITS file}

The main science output of the ALICE program is a multi-extension FITS file \citep{Pence2010} that gathers all the high-level information about the re-processed dataset. 

In an effort to facilitate high-level data exchange and to make exoplanet population statistical analyses easy, we developed a specific format for these data, and we propose that it becomes a standard for reduced products of high-contrast imaging data \citep{2014SPIE.9147E..51C}. We describe below the main choices we made for the Version 2 of this format and provide a detailed description of its content. 
We hereafter call this format the HCI-FITS file, for High-Contrast Imaging FITS file. It is inspired from the OI-FITS format, which is the standard for calibrated data exchange from optical interferometers \citep{Pauls2004,Pauls2005}. In addition to the products delivered by ALICE, the HCI-FITS format presented here has also been implemented in the Direct Imaging Virtual Archive (DIVA), another large scale database dedicated to high-contrast imaging data \citep{2017arXiv170305322V}.\footnote{\href{ http://cesam.lam.fr/diva/}{ http://cesam.lam.fr/diva/}}

\paragraph{A single FITS file} 

To enable high-level science analyses of high-contrast imaging data, several products are mandatory (e.g. the reduced image, detection limits, source detections). In order to prevent information loss when exchanging data, all the information must be gathered in a single file. The FITS file format offers the structure needed to achieve that, through the use of extensions. Extensions may contain different types of data, including images (\verb+IMAGE+ extension type) which is appropriate for the reduced images, sensitivity and SNR maps, and multi-dimension tables (\verb+BINTABLE+ extension type) which is appropriate for radial detection limits, characteristics of potentially detected sources, or general characteristics of the reduced products. Moreover, having a single file as reduced product makes databases more convenient to both implement and use.

\paragraph{A flexible standard}

We developed HCI-FITS format to be compatible with any type of dataset, regardless of the instrument, or observing mode, or processing method used to obtain the final reduced products. It can be used for both ground-based of space observation, for coronagraphic and saturated imaging, for broad-band imaging, integral-field spectroscopy, and polarimetric imaging. Depending on the observer/analyst's choice to present the reduced data, a HCI-FITS file may contain products for either a single image or an image cube. For example, this format supports all options between combining all reduced images from an Integral Field Spectrograph (IFS) in one broad-band image and keeping each reduced image separated in a spectral cube, while tracking specific image characteristics in all cases.

\paragraph{Structure of the HCI-FITS format}

We identified five main products that are necessary for a high-level use of reduced HCI data: The reduced images, the SNR maps, the sensitivity maps (or ``noise'' maps), the radial detection limits (or ``contrast curves''), and the characteristics of any detected point sources. The HCI-FITS format is thus composed of 6 extensions, one for each of these products, plus one which tracks the main characteristics of each image provided in the file. The extensions may appear in any order in the FITS file, but must have mandatory \verb+EXTNAME+ values to enable compatibility between files. The structure of the HCI-FITS format is provided in Table~\ref{tab:targetLevelFITS}. The SOURCE\_DETECTION extension is optional but must respect the specified format if present. This structure is not exclusive and may include additional data in other extensions (e.g. intermediate products such as the instrument PSF image). Reading software or codes should not presume the presence of such additional extensions.

The mandatory products enable analyses such as detection limit comparisons and astrophysical signal comparisons, but does not enable precise characterization of unreported signal. Such characterisation requires a forward modeling process \citep{Lagrange2010,Milli2012,Soummer2012,Pueyo2016} for which intermediate products are needed (instrument PSF image, raw data, eigen-images of the PSF library). Such detailed characterization is out of the scope of the HCI-FITS format use.

The DATA\_INFORMATION extension is critical to identify the characteristics of each reduced image in the file. It is the extension that makes this format compatible with any collection of high-contrast images. It is a \verb+BINTABLE+ extension that must be composed of 12 fields that track the field orientation, polarization state, epoch, and spectral information of the images. It must have as many rows as images provided in the file, and if several images are provided, the order must be the same in the DATA\_INFORMATION table as in the image cubes. We present in Table~\ref{tab:DataInformation} the structure of this table.

\begin{deluxetable*}{lcll}
\tabletypesize{\scriptsize}
\tablecaption{DATA\_INFORMATION extension \label{tab:DataInformation}}
\tablehead{
    \colhead{Label}&    \colhead{Type\tablenotemark{a}}& \colhead{Description}& \colhead{Units/Form}
}
\startdata
\hline
	Image\_Number	            & I	& Index of the image in the cube &  \\
	Orientation	                & D	& Sky-orientation of the image vertical (+Y) axis\\ 
	& & from North axis (East of North) & Degrees\\
	Combined\_Rotation\_Angle	& D	& Parallactic angle combined in the image & Degrees\\
	Number\_of\_Exposures	    & I	& Number of exposures combined in the image & \\
	Exposure\_Time	            & D	& Total exposure time combined in the image & Seconds\\
	Observation\_Start	        & D	& Mod. Julian Date at the start of the exposure & Days\\
	Observation\_End	        & D	& Mod. Julian Date at the end of the exposure & Days\\
	UT\_Midpoint\_Date\_of\_Observation    & A	& UT date at the image mid-point & YYYY-MM-DD\\
	UT\_Midpoint\_Time\_of\_Observation    & A	& UT time at the image mid-point & HH:MM:SS\\
	Wavelength                  & D	& Effective wavelength of the image & Angstroms\\
	Bandwidth                   & D	& Effective bandwidth of the image & Angstroms\\
	Polarization                & A	& Polarization state of the image & 
\enddata
\tablenotetext{a}{L = logical (8 bit), I = integer (16 bit), E = real (32 bit), D = double (64 bit), A = character string (160 bit)}
\end{deluxetable*}

The DETECTION\_LIMIT extension is also \verb+BINTABLE+ and reports the radial point source detection limits for each image present in the file. It must be composed of two mandatory fields reporting the separation from the star and the corresponding detection limit. The header of this extension must indicate the confidence level of the detection limit using the mandatory \verb+NSIGMA+ keyword. As the high-contrast imaging community is currently in the process of improving the definition of detection limits \citep{Mawet2014,JensenClem}, we note that the header of this extension may be further developed. The structure of this extension is presented in Table~\ref{tab:DetLimit}.

\begin{deluxetable}{lcl}
\tabletypesize{\scriptsize}
\tablecaption{DETECTION\_LIMIT extension \label{tab:DetLimit}}
\tablehead{
    \colhead{Label}&    \colhead{Type\tablenotemark{a}}& \colhead{Description}
}
\startdata
\hline
	Radius	            & D ($N_{im}$)	& Radial separation from the star\\
	Detection\_Limit    & D	($N_{im}$)  & Point source detection limit
\enddata
\tablenotetext{a}{L = logical (8 bit), I = integer (16 bit), E = real (32 bit), D = double (64 bit), A = character string (160 bit)}
\end{deluxetable} 

The third optional \verb+BINTABLE+ extension reports the characteristics of the point sources detected in the data. For each source it must indicate its astrometry, photometry, and SNR in each image provided in the file.

\paragraph{Specifics of the ALICE HCI-FITS files}

For the specific case of ALICE, the HCI-FITS files always contain the reduced data for each combined-roll and for the combination of all images, so the ALICE products present cubes of $N_{im}=N_{roll}+1$, where $N_{roll}$ is the number of spacecraft orientations used to observe the target.

As the ALICE data were reprocessed using the PCA-based KLIP algorithm using large PSF libraries, we provide in the REDUCED\_DATA header some specific keywords describing the reduction parameters we used for the dataset (See Table~\ref{tab:targetLevelKeywords}).

The images in the SENSITIVITY\_MAP extension are computed from the temporal variance of the residual speckle field through the PSF library. To do so, we reprocessed the reference images from the PSF library with the same parameters as the science images, and rotated-combined groups of them with the same numbers, weights, and angles as for the science combined images. We then compute covariance matrix of these  combined reference images, convolve it with a $1\lambda/D$ aperture, and compute its square-root to estimate the temporal speckle noise map per resolution element. The images in the SNR\_MAP extension are computed by convolving the reduced images from the REDUCED\_DATA extension with the same aperture, and dividing it with the images from the SENSITIVITY\_MAP. The tables provided in the DETECTION\_LIMIT extension are the radial averages of the images in the SENSITIVITY\_MAP extension, computed in 2-pixel wide annuli. They are normalized by the stellar flux converted to count/s (keyword \verb+STARFLUX+ in the primary header) to give a measure of the point source detection limit in terms of contrast to the star. We provide in the header of these three extensions keywords describing the parameters used to compute these metrics (see Table~\ref{tab:targetLevelKeywords}). {It is important to note that the sensitivity map and the detection limit do not inlcude the processing throughput.}

\begin{deluxetable}{@{\hspace{0.2em}}lcl@{\hspace{0.2em}}}
\tabletypesize{\scriptsize}
\tablecaption{ALICE keywords in the HCI-FITS file extension headers \label{tab:targetLevelKeywords}}
\tablehead{
 \colhead{Keyword}	& \colhead{Type\tablenotemark{a}}	& \colhead{Description}
}
\startdata
\hline
\multicolumn{3}{c}{REDUCED\_DATA}\\
\hline
EXTNAME     &A& Extension name\\
BUNIT       &A& Brightness units\\
REDALGO     &A& Reduction algorithm\\
REDSTRAT\tablenotemark{b}    &A& Strategy to build the PSF library\\
EXCLANG     &E& Exclusion angle for ADI strategy (deg)\\
TKL         &I& Num. of subtracted KL-modes w. KLIP\\
NPSFLIBR    &I& Number of images in the PSF library\\
\hline
\multicolumn{3}{c}{SNR\_MAP}\\
\hline
EXTNAME     &A& Extension name\\
BUNIT       &A& Brightness units\\
NOISEMET    &A& Method used for the detection limit\\
APERTRAD    &D& Aperture radius (pix)\\
SPATSCAL    &A& Spatial unit\\
\hline
\multicolumn{3}{c}{SENSITIVITY\_MAP}\\
\hline
EXTNAME     &A& Extension name\\
BUNIT       &A& Brightness units\\
NOISEMET    &A& Method used for the detection limit\\
NSIGMA	    &I& Detection limit confidence level (sigma)\\
APERTRAD    &D& Aperture radius (pix)\\
SPATSCAL    &A& Spatial unit\\
\hline
\multicolumn{3}{c}{DETECTION\_LIMIT}\\
\hline
EXTNAME     &A& Extension name\\
NOISEMET    &A& Method used for the detection limit\\
NSIGMA	    &I& Detection limit confidence level (sigma)\\
APERTRAD    &D& Aperture radius (pix)\\
SPATSCAL    &A& Spatial unit
\enddata
\tablenotetext{a}{L = logical (8 bit), I = integer (16 bit), E = real (32 bit), D = double (64 bit), A = character string (160 bit)}
\tablenotetext{b}{Two values may be found in ALICE-generated HCI-FITS files: ``RDI'' when all images of the target have been excluded from the PSF library, or ``ADI+RDI'' when the the PSF library also includes images of the target acquired at a complementary orientation of the spacecraft, respecting the EXCLANG exclusion angle.}
\end{deluxetable} 

The characteristics of the detected sources in extension SOURCE\_DETECTION are computed from a matched-filter process with a synthetic{, unocculted} NICMOS PSF, computed for the corresponding filter element with the TinyTIM software package \citep{Krist2011}. The source astrometry is determined with the position that maximizes the cross-correlation between the reduced images and the normalized synthetic PSF. The photometry of the source is retrieved with the maximum value of the cross-correlation, subtracted from the local background level, corrected from post-processing over-subtraction with analytical forward modeling \citep{Soummer2012}, and corrected from correlation losses between the synthetic and the real PSF by using photometric calibration data acquired on calibration white dwarfs. The contrast is computed by normalizing the photometry by the stellar flux converted in count/s (keyword \verb+STARFLUX+ in the primary header). Table~\ref{tab:SourceDet} provides a description of the SOURCE\_DETECTION extension.

\begin{deluxetable}{lcll}
\tabletypesize{\scriptsize}

\tablecaption{SOURCE\_DETECTION extension \label{tab:SourceDet}}
\tablehead{
\colhead{Label}& \colhead{Type\tablenotemark{a}}& \colhead{Description}& \colhead{Units}
}
\startdata
\hline
	Candidate	            & I 	& Index of the detected source                  & \\
	SNR                     & D	($N_{im}$)  & SNR of the source                     & \\
	dRA                     & D	($N_{im}$)  & Relative R.A. from the star           & Arcseconds\\
	err\_dRA                & D	($N_{im}$)  & Uncertainty on the dRA                & Arcseconds\\
	dDEC                    & D	($N_{im}$)  & Relative Dec. from star    & Arcseconds\\
	err\_dDEC               & D	($N_{im}$)  & Uncertainty on the dDEC               & Arcseconds\\
	Sep                     & D	($N_{im}$)  & Separation from the star              & Arcseconds\\
	err\_Sep                & D	($N_{im}$)  & Uncertainty on separation         & Arcseconds\\
	PA                      & D	($N_{im}$)  & Position Angle (east of north)        & Degrees\\
	err\_PA                 & D	($N_{im}$)  & Uncertainty on the PA                 & Degrees\\
	Flux\_cs                & D	($N_{im}$)  & Photometry                            & Counts/Sec\\
	err\_Flux\_cs           & D	($N_{im}$)  & Uncertainty on Flux\_cs               & Counts/Sec\\
	Flux\_mag               & D	($N_{im}$)  & Photometry                            & Mag\\
	err\_Flux\_mag          & D	($N_{im}$)  & Uncertainty on Flux\_mag              & Mag\\
	Flux\_Jy                & D	($N_{im}$)  & Photometry                            & Jansky\\
	err\_Flux\_Jy           & D	($N_{im}$)  & Uncertainty on Flux\_Jy               & Jansky\\
	Flux\_erg               & D	($N_{im}$)  & Photometry                            & erg/cm\textsuperscript{2}/s/A\\
	err\_Flux\_erg          & D	($N_{im}$)  & Uncertainty on Flux\_erg              & erg/cm\textsuperscript{2}/s/A\\
	Contrast                & D	($N_{im}$)  & Contrast from the star                & \\
	err\_Contrast           & D	($N_{im}$)  & Uncertainty on the Contrast           & 
\enddata
\tablenotetext{a}{L = logical (8 bit), I = integer (16 bit), E = real (32 bit), D = double (64 bit), A = character string (160 bit). In instances where these values are unknown, the value \textit{NaN} is used.}
\end{deluxetable}

The primary header of the ALICE HCI-FITS files is described in Table~\ref{tab:HCIFitsPrimaryHeader} of Appendix~\ref{app:mainfitsprimaryHead}. It gathers a selection of keywords useful at a science level, and come from the raw HST FITS header, LAPLACE program added keywords, and from our work.

\subsection{Data image products}
In addition to the main HCI-FITS file which provides science metrics for the combined products of a dataset, we also provide a ``Products'' folder gathering intermediate products for each image in the dataset. These files are complementary to the combined HCI-FITS products. While the purpose of main HCI-FITS file is to provide high-level science metrics to quantify the detection limits and detected sources in the dataset, the data image products are useful for diagnostic and astrophysical signal forward modeling.

For each exposure, we provide a multi-extension FITS file gathering the products described in Table~\ref{tab:imageLevelFITS}. 
Unlike the main HCI-FITS file, these products are specific to ALICE and their format is not compatible with all type of high-contrast science products.

The REDUCED\_IMAGE extension provides the reduced image computed by the ALICE pipeline using the KLIP algorithm. The image is not derotated and is presented with the same field orientation as the raw NICMOS image. The star is centered on pixel (41, 41) with pixel (1,1) at the bottom-left of the image. The header of the extension provides detailed information on the reduction parameters used to compute the image (see Table~\ref{tab:imageFileHeaders}).

The RAW\_IMAGE extension provides the raw NICMOS image calibrated by the LAPLACE program. The LAPLACE image has been cropped to a field of view smaller than the NICMOS full frame ($80\times80$ pixels {or $\sim6.1\times6.1$ arcsec}), and the star is centered on pixel (41, 41). The header of the extension lists the keywords provided in the raw HST file, as well as the position of the star center in the full NICMOS field of view (See Table~\ref{tab:imageFileHeaders}).

The REDUCTION\_ZONE extension provides a binary image of the reduction zone (pixels with 0 value were excluded from the reduction). In most case the reduction zone correspond to the full image except for a central mask of a few pixels radius. The parameters used to define the reduction zone are provided in the extension header (see Table~\ref{tab:imageFileHeaders}).

The EIGEN\_IMAGES extension provides the cube of the first principal components of the PSF library used for the PSF subtraction, truncated at the number of components actually used to reduce the data. This cube can be used to analytically compute the impact of the PSF subtraction process on an astrophysical source using forward modeling. 

The REF\_FILE\_NAMES extension provides a table with the file-name of the NICMOS images composing the PSF library used to reduce the image. The table also provides the position of the star center in these reference images in the full NICMOS field of view, around which they have been aligned and cropped to $80\times80$ pixels.

\begin{deluxetable*}{llll}
\tabletypesize{\scriptsize}
\tablecaption{Structure of the exposure-level Science Product FITS files
\label{tab:imageLevelFITS}}
\tablehead{
\colhead{Label}&\colhead{Type} &\colhead{Dimensions}& \colhead{Description}
}
\startdata
\hline
	REDUCED\_IMAGE	    & IMAGE		&$N_x \times N_y$				& Reduced data image\\
	RAW\_IMAGE		    & IMAGE		&$N_x \times N_y$				&Calibrated raw data image\\
	REDUCTION\_ZONE		&IMAGE		&$N_x \times N_y$				& Reduction zone\\
	EIGEN\_IMAGES 	    &IMAGE		&$N_x \times N_y \times N_{kl}$	& Cube of eigen-images used for KLIP subtraction\\
 	REF\_FILE\_NAMES	&BINTABLE	&$N_{ref}\times 5$					&List of the reference image filenames
\enddata
\end{deluxetable*}

\subsection{Preview Folder}
Finally, we also deliver for each dataset a ``Preview'' folder that contains PDF and CSV files of the content of each extension of the main HCI-FITS file. The PSF files show images of each frame of the REDUCED\_DATA extension (one version with the point source detections circled and one version without), of the first frame (North-combined frame) of the SNR\_MAP extension and of the DETECTION\_LIMIT extension. The CSV files show all the data contained in each BINTABLE extension (DATA\_INFORMATION, DETECTION\_LIMIT, SOURCE\_DETECTION).

\section{Conclusion}
We have presented the re-processed NICMOS data that we re-analyzed as part of the ALICE program. We deliver these science products to the community so that they may aid population studies through detection limits and substellar candidate identifications. We also presented the version 2 of the HCI-FITS format, a standard format for high-contrast imaging science products that can be used with any type of high-contrast imaging dataset. We hope this effort will help gather consistent datasets throughout the community.

\acknowledgments
This project was made possible by the Mikulski Archive for Space Telescopes (MAST) at STScI  which is operated by AURA, Inc. for NASA under contract NAS5-26555. Support was provided by NASA through grants HST-AR-12652.01 (PI: R. Soummer) and HST-GO-11136.09-A (PI: D. Golimowski), HST-GO-13855 (PI: E. Choquet), HST-GO-13331 (PI: L. Pueyo), and by STScI Director’s Discretionary Research funds. 
E.C. acknowledges support from NASA through Hubble Fellowship grant HF2-51355 awarded by STScI. Part of this research was carried out at the Jet Propulsion Laboratory, California Institute of Technology.
This research has made use of the SIMBAD database, operated at CDS, Strasbourg, France, and of NASA's Astrophysics Data System.\\

\appendix

\section{The NICMOS coronagraphic archive}

We report in Table~\ref{tab:nicmosPrograms} the comprehensive list of HST programs that used the coronagraphic mode of the NICMOS instrument for non-polarimetric observations. We also report the number of target observed per program, as well as the number of these which have been re-calibrated as part of the LAPLACE program, and re-analyzed as part of the ALICE program.

\begin{deluxetable*}{@{\hspace{.8em}}r@{\hspace{.8em}}rl@{}rr@{\hspace{.8em}}l@{\hspace{.8em}}p{0.61\linewidth}}
\tabletypesize{\scriptsize}
\tablecaption{Non-polarimetric NICMOS coronagraphic programs\label{tab:nicmosPrograms}}
\tablehead{
 \colhead{Prog.} &\colhead{Cy.} &\colhead{Type} &\colhead{LAPL\tablenotemark{a}}& \colhead{ALICE\tablenotemark{b}} 	&\colhead{PI}& \colhead{Title}}
\startdata
 \multicolumn{7}{c}{ERA 1}\\
 \hline
7038	&7	&	SM2/NIC	&0/3&0/3	&G. Schneider	&NICMOS Target Acquisition Test\\
7052	&7	&	SM2/NIC	&0/1&0/1	&G. Schneider	&NICMOS Coronographic Performance Verification\\
 7157 &7	&	SM2/NIC	&0/1&0/1	&G. Schneider	&NICMOS Optimum Coronagraphic Focus Determinaton\\
 7179 &7	&	GTO/NIC	&1/1&0/1	&B. Smith	&Search for Volcanic Activity on Encleadus\\
 7220 &7	&	GTO/NIC	&4/5&4/5	&R. Weymann	&Imaging of Quasar Host Galaxies\\
 7221 &7	&	GTO/NIC	&3/3&0/3	&J. Hill			&Imaging of Quasar Absorber Systems\\
 7226 &7	&	GTO/NIC	&44/44&39/44	&E. Becklin		&Search for Massive Jupiters\\
 7227 &7	&	GTO/NIC	&29/29&28/29	&G. Schneider	&A Search for Low Mass/Sub-Luminous Companions to M-Stars\\
 7233 &7	&	GTO/NIC	&18/23&17/23	&B. Smith	&Dust Disks around Main Sequence Stars\\
 7248 &7	&	GTO/NIC	&0/1&1/1	&B. Smith	&Spectroscopy and Polarimetry of the Beta Pictoris\\
 7329 &7	&	SNAP	&20/20&0/20	&M. Malkan	&The Nature of the Damped LyAlpha Absorbers -- A New Study of Young Galaxies\\
7418	&7	&	GO		&3/3&3/3	&D. Padgett	&NICMOS Imaging of Young Stellar Object Circumstellar Nebulosity\\
 7441 &7	&	GO		&0/6&0/6	&M. Brown	&A Search for Zodiacal Dust around Bright Nearby Stars\\
 7808 &7	&	ENG/NIC	&1/1&1/1	&G. Schneider	&NICMOS Coronagraphic Hole Location Test\\
 7828 &7	&	GO		&0/1&0/1	&J. Hollis		&Detection of the Infrared Jet in the R Aquarii Binary System\\
 7829 &7	&	GO		&6/6&5/6	&C. Johns-Krull	&Mapping H\_2 Emission Around T Tauri Stars\\
 7834 &7	&	GO		&7/7&6/7	&R. Rebolo		&A Search for Giant Planets Around Very Young Nearby Late-type Dwarfs\\
 7835 &7	&	GO		&8/14&7/14	&E. Rosenthal	&A Search for Superplanets Embedded in Beta Pic\&Vega-like Circumstellar Disks\\
 7857 &7	&	GO		&7/8&7/8	&A.-M. Lagrange	&Investigating the missing link between disks around Pre Main Sequence and Main Sequence stars\\
 7897 &7	&	SNAP	&20/20&17/20	&M. Clampin		&Probing planetary formation around main- sequence stars: A snapshot survey\\
 7924 &7	&	ENG/NIC	&1/1&1/1	&G. Schneider	&NICMOS Coronagraphic Hole Location Re-Test\\
 8079 &7	&	CAL/NIC	&2/2&2/2	&A. Schultz		&Mapping the Light Scatter in the NICMOS Coronagraphic Hole \{PSF Recovery Orbits\}\\
 \hline
 \multicolumn{7}{c}{ERA 2}\\
 \hline
8979	 &10		&SM3/NIC	&1/1&1/1	&G. Schneider	&NICMOS Optimum Coronagraphic Focus Determinaton\\
 8983 &10		&SM3/NIC	&2/2&2/2	&G. Schneider	&NICMOS Mode-2 Target Acquisition Test\\
 8984 &10		&SM3/NIC	&1/1&1/1	&G. Schneider	&NICMOS Coronagraphic Performance Assessment\\
 9693 &11		&CAL/NIC		&2/3&1/3	&G. Schneider	&NICMOS Coronagraphic Performance Assessment\\
 9834 &12		&GO			&7/7&6/7	&J. Debes		&Finding Planets in the Stellar Graveyard: A Faint Companion Search of White Dwarfs with NICMOS\\
 9845 &12		&GO/DD		&1/1&1/1	&M. Liu		&NICMOS Confirmation of a Young Planetary-Mass Companion\\
 10147 &13	&GO			&0/2&0/2	&M. Endl		&Detecting the elusive low mass companion around epsilon Indi\\
 10167 &13	&GO			&8/8&7/8	&A. Weinberger	&Imaging of Ices in Circumstellar Disks\\
 10176 &13	&GO			&111/111&108/111	&I. Song		&Coronagraphic Survey for Giant Planets Around Nearby Young Stars\\
 10177 &13	&GO			&56/56&54/56	&G. Schneider	&Solar Systems In Formation: A NICMOS Coronagraphic Survey of Protoplanetary and Debris Disks\\
10228 &13	&GO			&2/2&2/2	&P. Kalas		&Multi-color HST imaging of the GJ 803 debris disk\\
 10244 &13	&GO			&2/2&2/2	&M. Wyatt		&Coronagraphic imaging of Eta Corvus: a newly discovered debris disk at 18 pc\\
 10448 &13	&GO			&1/1&1/1	&A. Schultz		&NICMOS 2-gyro Coronagraphic Performance Assessment\\
 10464 &14	&ENG/NIC	&1/1&1/1	&A. Schultz		&NICMOS 2-gyro Coronagraphic Performance Assessment\\
 10487 &14	&GO			&11/11&9/11	&D. Ardila		&A Search for Debris Disks in the Coeval Beta Pictoris Moving Group\\
 10527 &14	&GO			&23/23&22/23	&D. Hines		&Imaging Scattered Light from Debris Disks Discovered by the Spitzer Space Telescope Around 20 Sun-like Stars\\
 10540 &14	&GO			&10/10&9/10	&A. Weinberger	&Imaging Nearby Dusty Disks\\
 10560 &14	&GO			&1/1&1/1	&J. Debes		&Confirming Planetary Candidates in the Stellar Graveyard with NICMOS\\
 10599 &14	&GO			&4/4&4/4	&P. Kalas		&Multi-color imaging of two 1 Gyr old debris disks within 20 pc of the Sun: Astrophysical mirrors of our Kuiper Belt\\
10847 &15	&GO			&0/1&0/1	&D. Hines		&Coronagraphic Polarimetry of HST-Resolved Debris Disks\\
 10849 &15	&GO			&23/23&23/23	&S. Metchev	&Imaging Scattered Light from Debris Disks Discovered by the Spitzer Space Telescope around 21 Sun-like Stars\\
 10852 &15	&GO			&3/6&2/6	&G. Schneider	&Coronagraphic Polarimetry with NICMOS: Dust grain evolution in T Tauri stars\\
 10854 &15	&GO			&0/6&0/6	&K. Stapelfeldt	&Coronagraphic Imaging of Bright New Spitzer Debris Disks II.\\
 10857 &15	&GO			&0/4&0/4	&A. Weinberger	&Are Organics Common in Outer Planetary Systems?\\
 10896 &15	&GO			&0/1&0/1	&P. Kalas		&An Efficient ACS Coronagraphic Survey for Debris Disks around Nearby Stars\\
 10897 &15	&GO			&0/2&1/2	&J.-F. Lestrade	&Coronagraphic imaging of the submillimeter debris disk of a 200Myr old M-dwarf\\
 11148 &16	&GO			&0/3&0/3	&J. Debes		&High Contrast Imaging of Dusty White Dwarfs\\
11155 &16	&GO			&0/8&5/8	&M. Perrin	&Dust Grain Evolution in Herbig Ae Stars: NICMOS Coronagraphic Imaging and Polarimetry\\
 11157 &16	&GO			&0/26&0/26	&J. Rhee		&NICMOS Imaging Survey of Dusty Debris Around Nearby Stars Across the Stellar Mass Spectrum
\enddata
\tablenotetext{a}{Number of targets in the program with at least one image re-calibrated by the LAPLACE program.}
\tablenotetext{b}{Number of targets in the program with at least one image re-processed by the ALICE program.}
\end{deluxetable*} 
\section{ALICE HCI-FITS file primary header}\label{app:mainfitsprimaryHead}

In the primary header of the HCI-FITS file, we selected keywords from the raw HST data keywords, and calibrated LAPLACE keywords that may be useful for a high-level science analysis of these data. We also added useful keywords specific to our work. The list of keyword present on the ALICE HCI-FITS file is presented in Table~\ref{tab:HCIFitsPrimaryHeader}.

\begin{deluxetable*}{lcll}
\tabletypesize{\scriptsize}
\tablecaption{Primary header of the HCI-FITS file for ALICE products\label{tab:HCIFitsPrimaryHeader}}
\tablehead{
 \colhead{Keyword}	&\colhead{Type\tablenotemark{a}} & \colhead{Description}& \colhead{Reference}}
\startdata
\hline
\multicolumn{4}{c}{File Information}\\
\hline
FILETYPE	&A&Type of data found in the file 	&This work\\
ORIGIN  	&A&FITS file originator				&FITS standard\\ 
DATE      	&A&Date this file was written (yyy-mm-dd) 	&FITS standard \\
\hline
\multicolumn{4}{c}{FITS File Structure}\\
\hline
DATATYPE 	&A&  Data type 	&This work\\
NEXTEND	    &I& Number of extensions	&This work\\
EXT[k]NAME	&A& Name of Extension [k]	&This work\\
EXT[k]TYPE	&A& Type of Extension [k]	&This work\\
\hline
\multicolumn{4}{c}{Data Structure and Description}\\
\hline
FRAMFORM	&A& Structure of each frame				&This work\\
FRAMENUM 	&I& Number of frames in the dataset     &This work\\
FRAME[k] 	&A& Description of frame [k]				&This work\\
\hline
\multicolumn{4}{c}{Program and Instrument Information}\\
\hline
TELESCOP	&A& Telescope used to acquire data				&HST keyword\\
INSTRUME	&A& Identifier for instrument used to acquire data	&HST keyword\\
PROPOSID	&I& PEP proposal identifier					&HST keyword\\
PR\_INV\_L	&A& Last name of principal investigator			&HST keyword\\
PR\_INV\_F	&A& First name of principal investigator			&HST keyword\\
CAMERA		&I& Camera in use (1, 2, or 3)					&HST keyword\\
FOCUS		&A& In-focus camera for this observation			&HST keyword\\
APERTURE	&A& Aperture in use							&HST keyword\\
FILTER		&A& Filter wheel element in beam during observation	&HST keyword\\
\hline
\multicolumn{4}{c}{Target Information}\\
\hline
TARGNAME	&A& Proposer's target name						&HST keyword\\
ALTNAME[k]\tablenotemark{b} &A& Alternative name \#[k] of the target	&SIMBAD \\
EQUINOX		&I& Equinox of celestial coord. system	&HST keyword\\
RA\_TARG 	&D& RA of target (deg) (J2000)  			&SIMBAD\\
DEC\_TARG 	&D& Declination of target (deg) (J2000) 	&SIMBAD\\
SC\_EMAJ	&D& Sky coord. error ellipse major axis	&SIMBAD\\
SC\_EMIN	&D& Sky coord. error ellipse minor axis	&SIMBAD\\
SC\_EPA		&D& Sky coord. error ellipse position angle	&SIMBAD\\
SC\_BIB 	&A& Sky coord. bibliography code 		&SIMBAD\\
PARALLAX\tablenotemark{c} 	&D& Parallax for target found (mas) 		&SIMBAD\\
PAR\_ERR\tablenotemark{c} 	&D& Parallax mean error 				&SIMBAD\\
PAR\_BIB\tablenotemark{c} 	&A& Parallax bibliography code 			&SIMBAD\\
PROPRA\tablenotemark{c} 	&D& Proper motion (RA) of target (mas/yr) 	&SIMBAD\\
PROPDEC\tablenotemark{c} 	&D& Proper motion (Dec) of target (mas/yr) &SIMBAD\\
PM\_EMAJ\tablenotemark{c} 	&D& Proper motion error ellipse major axis &SIMBAD\\
PM\_EMIN\tablenotemark{c} 	&D& Proper motion error ellipse minor axis &SIMBAD\\
PM\_EPA\tablenotemark{c} 	&D& Proper motion error ellipse position angle &SIMBAD\\
PM\_BIB\tablenotemark{c} 	&A& Proper motion bibliography code		&SIMBAD\\
\hline
\multicolumn{4}{c}{Information on other astrophysical sources}\\
\hline
CANDNUM &I& Number of point source detections	&This work\\
DISKDET &L& Disk detected in this dataset		&This work\\
\hline
\multicolumn{4}{c}{Photometric Information}\\
\hline
ADCGAIN		&D& Analog-digital conversion gain (electron/DN)&HST keyword\\
STARFLUX    &D& Star flux computed from synphot (count/s) & This work\\
J\_2MASS	&D& Target J band magnitude (2MASS catalog)	&LAPLACE keyword\\
JE\_2MASS	&D& 2MASS J magnitude uncertainty			&LAPLACE keyword\\
H\_2MASS	&D& Target H band magnitude (2MASS catalog)	&LAPLACE keyword\\
HE\_2MASS	&D& 2MASS H magnitude uncertainty			&LAPLACE keyword\\
K\_2MASS	&D& Target K band magnitude (2MASS catalog)	&LAPLACE keyword\\
KE\_2MASS	&D& 2MASS K magnitude uncertainty			&LAPLACE keyword\\
F160W\_EF	&L& K band SED flux density excess flag (Y,N)	&LAPLACE keyword\\
F160W\_JY	&D& Filterband target flux density estimate (Jy)	&LAPLACE keyword\\
PHOTMODE	&A& LAPLACE keyword\\
PHOTFLAM	&D& Inverse sensitivity (ergs/cm**2/Angstrom/DN)	&HST keyword\\
PHOTFNU	    &D& Inverse sensitivity (JY*sec/DN)	&HST keyword\\
PHOTZPT		&D& ST magnitude system zero point (mag)	&HST keyword\\
PHOTPLAM	&D& Pivot wavelength of the photmode (Angstrom)	&HST keyword\\
PHOTBW		&D& RMS bandwidth of the photmode (Angstrom)	&HST keyword\\
\hline
\multicolumn{4}{c}{Astrometric Information}\\
\hline
PIXSCALE 	&D&  Pixel scale (arcsec)&
\enddata

\tablenotetext{a}{L = logical (8 bit), I = integer (16 bit), E = real (32 bit), D = double (64 bit), A = character string (160 bit)}
\tablenotetext{b}{For this dataset delivery, we only provide ALTNAME1 -- the resolved SIMBAD name.}
\tablenotetext{c}{When the information is available.}
\end{deluxetable*}

\section{Data image Fits file headers}
In Table ~\ref{tab:imageFileHeaders} we describe the header of the Data image FITS file provided in the ``Products'' folder. Most of the keywords in the primary header come from the raw NICMOS FITS file primary header, and we only describe here the keywords that we modified or added. Similarly, the RAW\_IMAGE extension header corresponds to the SCI extension header in the raw NICMOS file, and we describe her the added keywords.

\begin{deluxetable*}{lcl}
\tabletypesize{\scriptsize}
\tablecaption{Header of the Data image FITS file\label{tab:imageFileHeaders}}
\tablehead{
 \colhead{Keyword} &\colhead{Type\tablenotemark{a}} 	& \colhead{Description}
}
\startdata
\multicolumn{3}{c}{Added/edited keywords in the Primary Header}\\
\hline
 FILETYPE   &A& Type of data found in data file\\
 ORIGIN     &A& FITS file originator\\
 DATE       &A& Date this file was written (yyyy-mm-dd)\\
 DATATYPE   &A& Data type\\
 NEXTEND    &I& Number of standard extensions\\
 EXT[k]NAME	&A& Name of Extension [k]	\\
 EXT[k]TYPE	&A& Type of Extension [k]	\\
 FRAMFORM   &A& Structure of the frame\\ 
ALTNAME1    &A& Simbad name for target\\
PIXSCALX    &E& Pixel scale in X direction\\
PIXSCALY    &E& Pixel scale in Y direction\\
\hline
\multicolumn{3}{c}{Keywords in the REDUCED\_IMAGE extension header}\\
\hline
EXTNAME     &A& Extension name\\
BUNIT       &A& Brightness units\\
REDALGO     &A& Reduction algorithm\\
REDSTRAT    &A& Strategy to build the PSF library\\
EXCLANG     &E& Exclusion angle for ADI-type strategy (deg)\\
TKL         &I& Number of eigenimages used for KLIP-type subtraction\\
NPSFLIBR    &I& Number of images in the PSF library \\
STARCENX    &E& Star center X position in the full NICMOS frame (pix)\\
STARCENY    &E& Star center Y position in the full NICMOS frame (pix)\\
\hline
\multicolumn{3}{c}{Added/edited keywords in the RAW\_IMAGE extension header}\\
\hline
EXTNAME &A& Extension name\\
STARCENX &E& Star center X position in the full NICMOS frame (pix)\\
STARCENY &E& Star center Y position in the full NICMOS frame (pix)\\
\hline
\multicolumn{3}{c}{Keywords in the REDUCTION\_ZONE extension header}\\
\hline
EXTNAME &A&  Extension name\\
BUNIT &A& Brightness units\\
ZONECPA &I& Zone center PA to North (deg) \\
ZONEDPA &I& Zone delta PA (deg) \\
ZONERIN &I& Zone inner radius (pix) \\
ZONEROU &I& Zone outer radius (pix) \\
\hline
\multicolumn{3}{c}{Keywords in the EIGEN\_IMAGES extension header}\\
\hline
EXTNAME &A& Extension name\\
REDALGO &A&  Reduction algorithm\\
REDSTRAT &A& Strategy to build the PSF library\\
EXCLANG &E& Exclusion angle for ADI-type strategy (deg) \\
NPSFLIBR &I& Number of images in the PSF library
\enddata
\tablenotetext{a}{L = logical (8 bit), I = integer (16 bit), E = real (32 bit), D = double (64 bit), A = character string (160 bit)}
\end{deluxetable*}

\bibliographystyle{apj.bst} 
\bibliography{biblio-disk-planet}

\end{document}